\documentclass[prl,twocolumn,showpacs,amsmath,amssymb]{revtex4-1}

\usepackage{graphicx}
\usepackage{subfigure}
\usepackage{dcolumn}
\usepackage{array}
\usepackage{bm}
\usepackage[hang]{footmisc}

\usepackage[usenames,dvipsnames]{color}

\begin{document}

\title{\bf{Antiferroelectricity in thin film ZrO$_2$ from first principles} \\[11pt] } \author{Sebastian E. Reyes-Lillo, Kevin F. Garrity and Karin M. Rabe}
\affiliation{Department of Physics and Astronomy, Rutgers University, Piscataway, NJ 08854-8019}\date{\today}

\begin{abstract}
Density functional calculations are performed to investigate the experimentally-reported field-induced phase transition in thin-film ZrO$_2$ (J. M$\ddot{u}$ller \emph{et al.}, Nano. Lett. {12}, 4318). We find a small energy difference of~$\sim$~1~meV/f.u. between the nonpolar tetragonal and polar orthorhombic structures, characteristic of antiferroelectricity. The requisite first-order transition between the two phases, which atypically for antiferroelectrics have a group-subgroup relation, results from coupling to other zone-boundary modes, as we show with a Landau-Devonshire model. Tetragonal ZrO$_2$ is thus established as a previously unrecognized lead-free antiferroelectric with excellent dielectric properties and compatibility with silicon. In addition, we demonstrate that a ferroelectric phase of ZrO$_2$ can be stabilized through epitaxial strain, and suggest an alternative stabilization mechanism through continuous substitution of Zr by Hf. 
\end{abstract}

\pacs{
77.84.-s
81.05.Zx, 
77.65.Bn, 
}

\maketitle

\marginparwidth 2.5in
\marginparsep 0.5in
\def\jab#1{\marginpar{\small JAB: #1}}
\def\jwb#1{\marginpar{\small JWB: #1}}
\def\hit#1{\marginpar{\small HIT: #1}}
\def\amr#1{\marginpar{\small AMR: #1}}
\def\kfg#1{\marginpar{\small KFG: #1}}
\def\scr{\scriptsize}

\begin{figure}[t]
\includegraphics[scale=0.45]{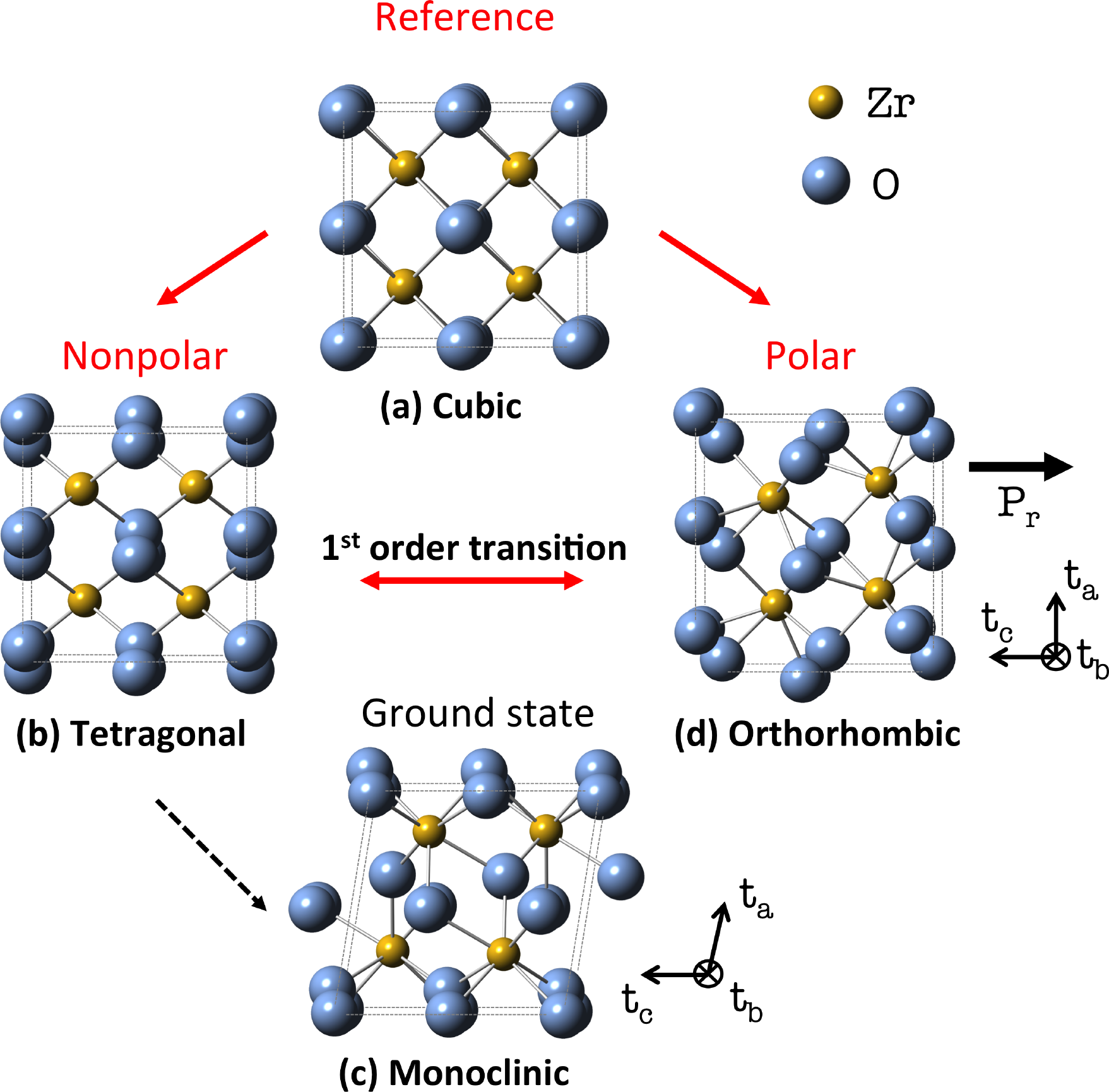}
\caption{\label{fig:zro2-phases} Experimentally reported phases of ZrO$_2$: (a) Cubic (\textsl{Fm$\bar{3}$m}), (b) Tetragonal (\textsl{P4$_2$/nmc}), (c) Monoclinic (\textsl{P2$_1$/c}) and (d) Orthorhombic (\textsl{Pca2$_1$}).}
\end{figure}

Zirconia (ZrO$_2$) is a high-k dielectric~\cite{Bohr2007}, chemically and structurally similar to HfO$_2$, and likewise is a candidate for dynamic random access memory (DRAM) applications~\cite{Kittl2009, Hwang2010} and complementary metal-oxide-semiconductor (CMOS) devices~\cite{Choi2011, Panda2013}. Bulk ZrO$_2$ has a high-symmetry cubic (\textsl{Fm$\bar{3}$m}) structure (Fig. \ref{fig:zro2-phases}(a)) above~2400~K, and a tetragonal (\textsl{P4$_2$/nmc}) structure (Fig. \ref{fig:zro2-phases}(b)) between~2400~K and~1200~K~\cite{Ruh1968}. The tetragonal structure is related to the cubic structure by freezing in an unstable $X_2^-$ mode~\cite{Mirgorodsky1995} and is nonpolar. Below~1200~K, ZrO$_2$ is monoclinic (\textsl{P2$_1$/c}) (Fig. \ref{fig:zro2-phases}(c)). The first-order transition from the tetragonal phase to the monoclinic phase changes the coordination number of Zr from 8 to 7 and increases the volume by~$\sim$~5~\%. 
\medskip

In light of the extensive research which has been conducted over the past fifty years on this relatively simple dielectric, the recent report of antiferroelectric-like double-hysteresis loops in thin film ZrO$_2$~\cite{Muller2012} at first seems rather surprising. In thin film ZrO$_2$, the tetragonal-monoclinic transition temperature is suppressed and the structure is tetragonal at room temperature~\cite{Garvie1978, Kim2004, Keun2008}; in contrast, thin film HfO$_2$ is monoclinic at room temperature and exhibits simple dielectric behavior. The field-induced polar phase in ZrO$_2$, which appears above a critical field on the order of 2~MV/cm, is isostructural with the ferroelectric phases that have been observed in thin films of HfO$_2$ doped with Al~\cite{SMueller2012a}, Y~\cite{Muller2011a}, Gd~\cite{SMueller2012b}, Si~\cite{Boscke2011a, Boscke2011b} and Sr~\cite{Schenk2013}, as well as in (Hf$_{1/2}$Zr$_{1/2}$)O$_2$ thin films~\cite{Muller2011b, Park2013}. The structure of the polar phase is orthorhombic (\textsl{Pca2$_1$})~\cite{Kisi1989} and corresponds to a distortion of the high-symmetry cubic structure, as depicted in Fig.~\ref{fig:zro2-phases}. 
\medskip

Antiferroelectrics have recently been the subject of increasing interest~\cite{Rabe2013}. The characteristic electric-field-induced transition from a nonpolar to a strongly polar phase is the source of functional properties and promising technological applications. Non-linear strain and dielectric responses due to the phase switching are useful for transducers and electro-optic applications~\cite{Berlincourt1966, Zhang2009}. The shape of the double hysteresis loop suggests applications in high-energy storage capacitors~\cite{Jaffe1961, Love1990}. In addition, an electro-caloric effect can be observed in systems with a large entropy change between the two phases~\cite{Mischenko2006}. While most attention has focused on PbZrO$_3$ and related perovskites~\cite{Tan2011}, a recent theoretical materials design search~\cite{Bennett2013} suggested that there are many more antiferroelectric compounds to be discovered.
\medskip

In this work, we use first-principles calculations to provide clear evidence that the tetragonal phase of ZrO$_2$ is a previously unrecognized antiferroelectric material, and that the behavior observed in thin films is intrinsic. We find a remarkably small energy difference of~$\sim$~1~meV per formula unit (meV/f.u.) between the nonpolar tetragonal and polar orthorhombic structures, which is a key characteristic of antiferroelectricity~\cite{Singh1995, Johannes2005, Reyes-Lillo2013, Lasave2007}. We show with a polynomial expansion of the energy that the requisite first-order transition between nonpolar and polar phases, which in the present case, atypically for antiferroelectrics, have a group-subgroup relation, results from coupling to other zone-boundary modes. This novel mechanism for antiferroelectricity provides a basis for broader searches for antiferroelectric materials with optimal functional properties. In addition, we demonstrate that the polar phase of ZrO$_2$ can be stabilized through epitaxial strain, and we discuss the possibility of ferroelectricity in the solid solution (Zr$_{1-x}$Hf$_{x}$)O$_2$.
\medskip

Density-functional theory (DFT) calculations are performed using version 7.4.1~of~\texttt{ABINIT}~\cite{Gonze2009} and the local-density approximation~(LDA).  We use a plane-wave energy cutoff of~680~eV, and a~$4 \times 4 \times 4$ Monkhorst-Pack sampling of the Brillouin zone~\cite{Monkhorst1976} for all structural optimizations. Polarization was calculated on a~$10 \times 10 \times 10$ grid using the modern theory of polarization~\cite{King-Smith1993} as implemented in \texttt{ABINIT}. We used norm-conserving pseudopotentials from the Bennett-Rappe library~\cite{Bennett2012} with reference configurations: \textsl{Zr}~([\textsl{Kr}]4d$^0$5s$^0$), \textsl{Hf}~([\textsl{Xe}]4f$^{14}$5d$^{0.5}$6s$^0$) and \textsl{O}~([\textsl{He}]2s$^2$2$p^4$), generated by the \texttt{OPIUM} code~\cite{opium}. 
\medskip

Table~\ref{table:polar-phases-zro2} reports relaxed bulk energies and lattice constants for the experimentally observed bulk phases of ZrO$_2$. Our results are in good agreement with previous first principles results~\cite{Jansen1991, Lowther1999, Rignanese2001, Zhao2002a}. The monoclinic phase is the lowest energy structure, consistent with experiments. Experimental lattice constants~\cite{Ding2012, Jin2003, Kisi1989, Jovalekic2008} are underestimated by~1~\%, typical for LDA. The spontaneous polarization of the orthorhombic structure was found to be~P$_r$~$\sim$~58~$\mu$C/cm$^2$. 
\medskip

\begin{table}[t]
\caption{Energy $\Delta E$ (meV/f.u.), volume expansion $\Delta V/V$ ($\%$) and lattice parameters (latt.) in~\AA~($\beta$: monoclinic angle) of the experimental phases of ZrO$_2$.}
\begin{ruledtabular}
\begin{tabular}{lcccccc}
Phase & $\Delta E$ & $\Delta V/V$ & latt. & This work  & Prev. work$^a$  & Exp.$^b$   \\ \hline
\begin{tabular}{l}  \textsl{Fm$\bar{3}$m}  \\ \end{tabular}
&\begin{tabular}{c}     0\\  \end{tabular}
&\begin{tabular}{c}     0\\  \end{tabular}
&\begin{tabular}{c}     a$_0$\\  \end{tabular}
&\begin{tabular}{c}   5.03\\  \end{tabular}
&\begin{tabular}{c}   5.04\\  \end{tabular}
&\begin{tabular}{c}   5.03\\  \end{tabular}   \\ \hline
\begin{tabular}{l} \textsl{P4$_2$/nmc}\\  \\ \end{tabular}
&\begin{tabular}{c}       -47.8\\ \\ \end{tabular} 
&\begin{tabular}{c}       2.0  \\ \\ \end{tabular} 
&\begin{tabular}{c}       a    \\ c\\ \end{tabular} 
&\begin{tabular}{c}        5.04\\ 5.12\\ \end{tabular} 
&\begin{tabular}{c}        5.03\\ 5.10\\ \end{tabular} 
&\begin{tabular}{c}        5.09\\ 5.19\\ \end{tabular}    \\ \hline
\begin{tabular}{l} \textsl{Pca2$_1$}\\  \\ \\\end{tabular}
&\begin{tabular}{c} -48.2\\  \\ \\  \end{tabular}
&\begin{tabular}{c}   3.6\\  \\ \\  \end{tabular}
&\begin{tabular}{c}   a\\  b\\ c\\  \end{tabular}
&\begin{tabular}{c} 5.22\\ 5.02\\ 5.04\\  \end{tabular}
&\begin{tabular}{c} 5.26\\ 5.07\\ 5.08\\  \end{tabular}
&\begin{tabular}{c} 5.26\\ 5.07\\ 5.08\\  \end{tabular}   \\ \hline
\begin{tabular}{l} \textsl{P2$_1$/c}\\ \\ \\ \\ \end{tabular}
&\begin{tabular}{c} -82.0\\   \\  \\ \\ \end{tabular}
&\begin{tabular}{c}  7.2 \\   \\  \\ \\ \end{tabular}
&\begin{tabular}{c}     a\\  b\\ c\\ $\beta$\\ \end{tabular}
&\begin{tabular}{c} 5.09\\ 5.20\\ 5.24\\ 99.39\\ \end{tabular}
&\begin{tabular}{c} 5.11\\ 5.17\\ 5.27\\ 99.21\\ \end{tabular}
&\begin{tabular}{c} 5.14\\ 5.20\\ 5.31\\ 99.17\\ \end{tabular} 
\end{tabular}
\end{ruledtabular}
$^a$ Refs.~\cite{Zhao2002a, Lowther1999} \\
$^b$ Refs.~\cite{Ding2012, Jin2003, Kisi1989, Jovalekic2008}
\label{table:polar-phases-zro2}
\end{table}

\begin{table}[t]
\caption{Wave-vector ($q$) in reciprocal lattice units (2$\pi$/a$_0$) and eigenvector of the structural modes contained in the \textsl{Pca2$_1$} structure, specified by the independent non zero displacements. With Zr at the origin, O$_1$ at a$_0$/4(111) and O$_2$ at a$_0$/4(1$\bar{1}$1). $Q_{th}$ and $Q_{ex}$ are the computed and experimental mode amplitudes, defined as described in the text.}
\begin{ruledtabular}
\begin{tabular}{l|c|c|c|c|c}
   & $\Gamma_4^-$ & $X_2^-$     &

\begin{tabular}{ccc}

\begin{tabular}{c} $X_{5,x}^+$\\ \end{tabular}
&\begin{tabular}{c} $X_{5,y}^+$\\  \end{tabular} 
&\begin{tabular}{c} $X_{5,z}^+$\\  \end{tabular} \\
\end{tabular}

& $X_5^-$ & $X_3^-$ \\

$q$    

&$\hat{0}$
&$\hat{x}$        
&\begin{tabular}{ccc}

\begin{tabular}{l}  $\hat{z}$\\ \end{tabular}
&\begin{tabular}{c} $\hat{x}$\\  \end{tabular} 
&\begin{tabular}{r} $\hat{y}$\\  \end{tabular} \\
\end{tabular}

& $\hat{z}$

&$\hat{y}$  \\ \hline
\begin{tabular}{l}  Zr\\ O$_1$\\ O$_2$\\ \end{tabular}
&\begin{tabular}{c} \phantom{$-$}u$_z$\\ $-$u$_z$\\ $-$u$_z$\\ \end{tabular}
&\begin{tabular}{c} \phantom{$-$}0\\ $-$u$_x$\\ \phantom{$-$}u$_x$\\ \end{tabular}

&\begin{tabular}{ccc}
\begin{tabular}{c} \phantom{$-$}0\\ $-$u$_x$\\ $-$u$_x$\\ \end{tabular}
&\begin{tabular}{c} \phantom{$-$}0\\ \phantom{$-$}u$_y$\\ \phantom{$-$}u$_y$\\ \end{tabular}
&\begin{tabular}{c} \phantom{$-$}0\\ \phantom{$-$}u$_z$\\ $-$u$_z$\\ \end{tabular} \\
\end{tabular}

&\begin{tabular}{cc}
\begin{tabular}{c} \phantom{$-$}0\\ \phantom{$-$}u$_y$\\ $-$u$_y$\\ \end{tabular}
&\begin{tabular}{c} \phantom{$-$}u$_x$\\ \phantom{$-$}0\\ \phantom{$-$}0\\  \end{tabular} \\
\end{tabular}
&\begin{tabular}{c} \phantom{$-$}u$_y$\\ \phantom{$-$}0\\ \phantom{$-$}0\\ \end{tabular} \\ \hline
$Q_{th}$ & 0.259 & 0.499 

&\begin{tabular}{ccc}

\begin{tabular}{c} 0.377\\ \end{tabular}
&\begin{tabular}{c} 0.394\\  \end{tabular} 
&\begin{tabular}{c} 0.392\\  \end{tabular} \\
\end{tabular}

&\begin{tabular}{ccc}

\begin{tabular}{c} 0.119\\ \end{tabular}
&\begin{tabular}{c} 0.159\\  \end{tabular} \\
\end{tabular}

& 0.089 \\

$Q_{ex}$ & 0.233 & 0.491 

&\begin{tabular}{ccc}

\begin{tabular}{c} 0.335\\ \end{tabular}
&\begin{tabular}{c} 0.376\\  \end{tabular} 
&\begin{tabular}{c} 0.381\\  \end{tabular} \\
\end{tabular}

&\begin{tabular}{ccc}

\begin{tabular}{c} 0.111\\ \end{tabular}
&\begin{tabular}{c} 0.158\\  \end{tabular} \\
\end{tabular}

& 0.086 \\
\end{tabular}
\end{ruledtabular}
\label{table:orthorhombic-modes}
\end{table}

We begin by discussing the energy-lowering distortions of the cubic phase of ZrO$_2$. The zone-boundary $X_2^-$ mode corresponds to an antipolar displacement of oxygen atoms in the $\hat{x}$ direction. It is the single instability exhibited in the first-principles phonon dispersion of the high-symmetry cubic structure~\cite{Parlinski1997} and, as mentioned above, leads to the tetragonal \textsl{P4$_2$/nmc} phase with amplitude~0.319~\AA~and 4-fold axis along $\hat{x}$. We find that the tetragonal structure is a local minimum, with no unstable modes. Therefore, even though the symmetry analysis shows that the transition from the $\hat{z}$-polarized tetragonal structure to the orthorhombic \textsl{Pca2$_1$} phase is allowed to be second order, the field-induced transition is expected to be first order.
\medskip

The distortion relating the tetragonal \textsl{P4$_2$/nmc} structure to the polar orthorhombic \textsl{Pca2$_1$} phase can be decomposed into symmetry-adapted modes of the high-symmetry cubic \textsl{Fm$\bar{3}$m} structure. Mode amplitudes are computed with respect to eigenvectors normalized to 1~\AA. Table~\ref{table:orthorhombic-modes} shows excellent agreement between first principles ($Q_{th}$) and experimental ($Q_{ex}$) values of the mode amplitudes. There are six nonzero modes in addition to the $X_2^-$ mode: $\Gamma_4^-$, $X_{5,x}^+$, $X_{5,y}^+$, $X_{5,z}^+$, $X_5^-$ and $X_3^-$~\cite{fm-3m-irrep}. The zone-center $\Gamma_4^-$ mode corresponds to a polar displacement of the zirconium atoms relative to the oxygen atoms in the $\hat{z}$ direction. The three $X_5^+$ modes involve antipolar displacements of planes of oxygen atoms. The amplitudes of the $X_5^-$ and $X_3^-$ modes are relatively small.
\medskip

\begin{figure}[t]
\includegraphics[scale=0.5]{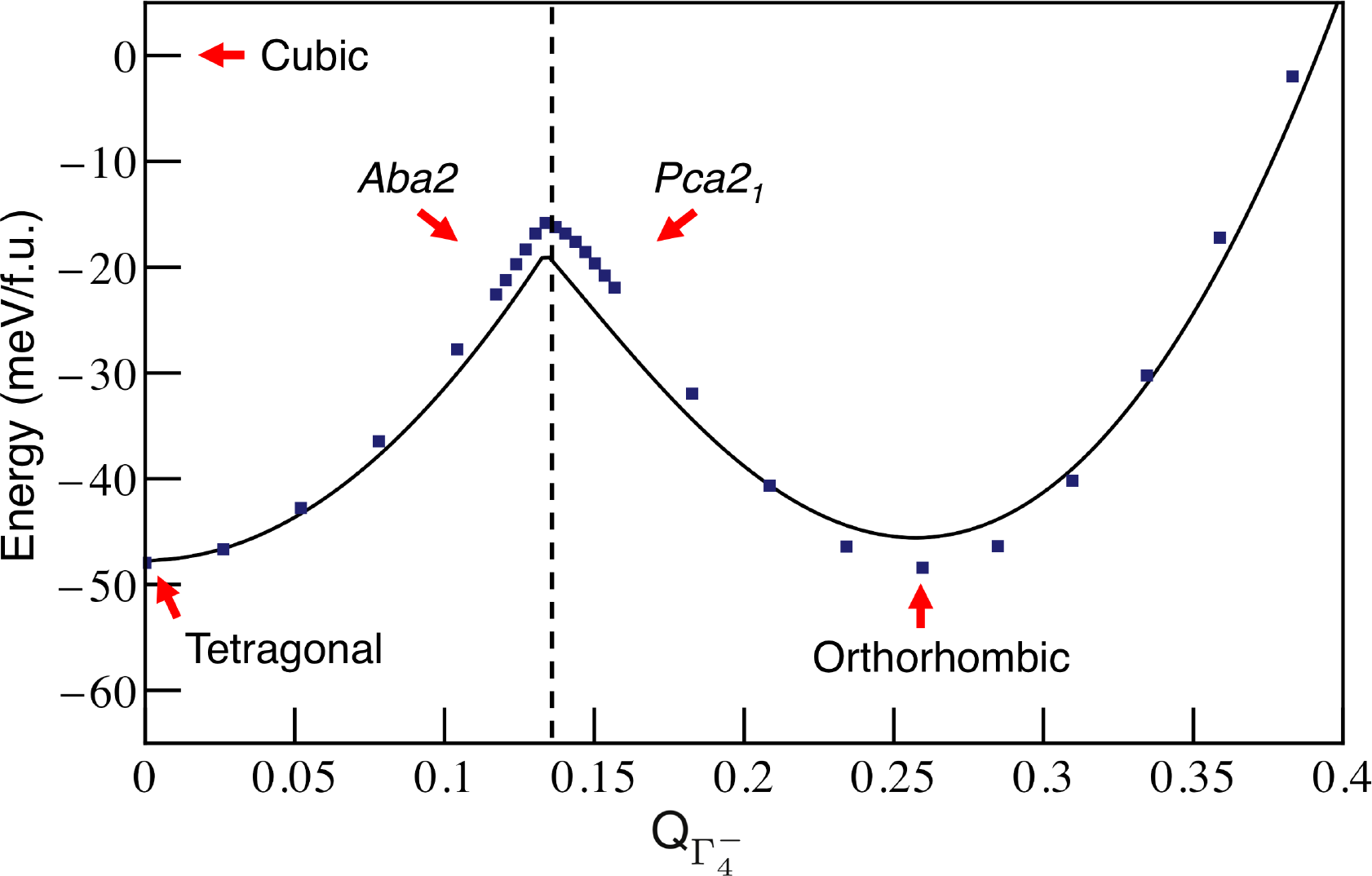}
\caption{\label{fig:zro2-barrier} First-principles calculations (squares) and Landau-Devonshire model (solid line) of the energy profile between tetragonal and orthorhombic phases.}
\end{figure}

We explore the field-induced tetragonal-to-orthorhombic transition by performing relaxations of the structures obtained by linear interpolation of the atomic positions, holding fixed the value of the polar mode amplitude $Q_{\Gamma_4^-}$. Fig.~\ref{fig:zro2-barrier} shows the energy of the polar structure as a function of $Q_{\Gamma_4^-}$. For small values of~$Q_{\Gamma_4^-}$, the energy of the system increases, as expected for freezing in a stable mode. The induced polar structure has an orthorhombic space group \textsl{Aba2}, different from that of the polar orthorhombic \textsl{Pca2$_1$} phase, and relaxes back to the tetragonal phase if the constraint is removed. At~$Q_{\Gamma_4^-}$~$\sim$~0.134~\AA, we find an energy cusp with magnitude $\delta$E~$\sim$~35~meV/f.u., which separates the \textsl{Aba2} structure with Q$_{X_{5,y}^+}$$\neq$0 and Q$_{X_{5,x,z}^+}$=0, from the \textsl{Pca2$_1$} phase with Q$_{X_{5,x,y,z}^+}$$\neq$0~\cite{pca21-variant}. Above this threshold value, the system volume expands ($\sim$~1.5~\%), and the structure relaxes to the polar orthorhombic \textsl{Pca2$_1$} phase if the constraint is removed. Assuming that the critical field E$_C$ required to overcome the energy barrier is given by E$_C$~$\sim$~ $\delta$E / V$_C$ P$_C$ (with P$_C$~=~36~$\mu$C/cm$^2$ and V$_C$=32.9~\AA$^3$/f.u., the first principles values of polarization and volume at the barrier), we estimate the critical field as E$_C$~$\sim$~4.7~MV/cm.
\medskip

In order to characterize this unusual energy surface, we describe the energetics of a single orthorhombic \textsl{Pca2$_1$} variant based on an expansion around the minimum-energy tetragonal \textsl{P4$_2$/nmc} structure using modes of the high-symmetry \textsl{Fm$\bar{3}$m} structure as a basis set:
\begin{align*}
E_T(Q_{\Gamma_4^-},Q_{X_{5,a,b,c}^+})&=E^0_{T}+a_1 Q^2_{\Gamma_4^-} +a_2 Q_{\Gamma_4^-} Q_{X_{5,y}^+}\\
                                     &+S(Q_{X_{5,x}^+},Q_{X_{5,y}^+},Q_{X_{5,z}^+}) \quad (1),
\end{align*}
\noindent where $S$ is a function to be determined below. The invariance of the term $a_2 Q_{\Gamma_4^-} Q_{X_{5,y}^+}$ originates in the fact that the expansion is around a structure of lower symmetry than the structure used to construct and label the modes (for details, see
Supplemental Material~\cite{supp-mat}).
\medskip 

\begin{figure*}[t]
\includegraphics[scale=0.64]{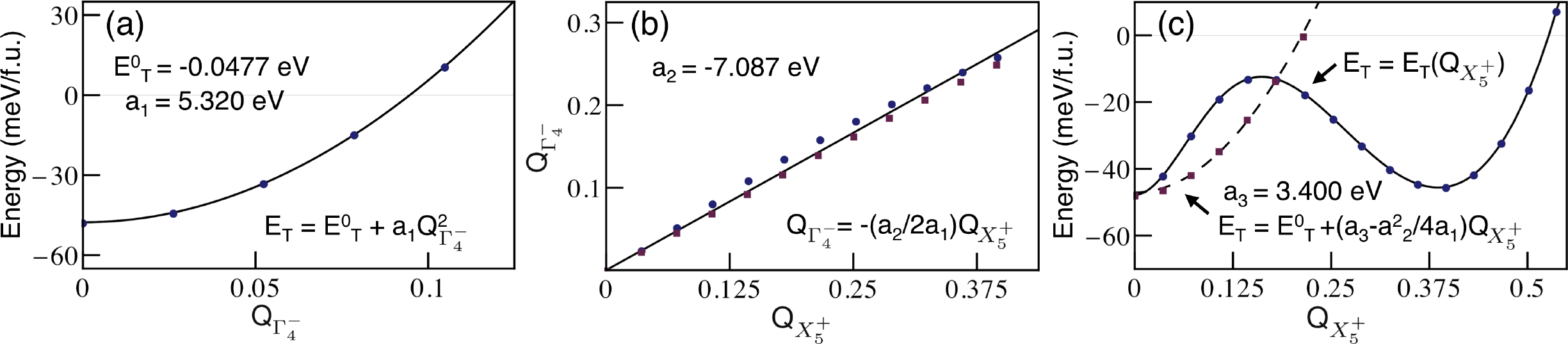}
\caption{\label{fig:zro2-fitting} First-principles calculations (circles and squares) and model fitting (solid and dashed lines). (a) Energy of the polar \textsl{Aba2} structure as a function of $Q_{\Gamma^-_4}$ with $Q_{X_{5,y}^+}$=0.  (b) $Q_{\Gamma^-_4}$ as a function of $Q_{X_{5,x,y,z}^+}$=$Q_{X^+_5}$ (circles) and $Q_{X_{5,y}^+}$=$Q_{X^+_5}$ (squares) (c) Energy of the polar \textsl{Pca2$_1$} structure as a function of $Q_{X_{5,x,y,z}^+}$=$Q_{X^+_5}$ (circles) and $Q_{X_{5,y}^+}$=$Q_{X^+_5}$ (squares).}
\end{figure*}

The values of the coefficients are obtained by fitting the model to first principles calculations. By freezing in $Q_{\Gamma_4^-}$ in the tetragonal phase, and relaxing the structure with $Q_{X^+_{5,y}}$ constrained to zero, we extract the values of $E^0_{T}$ and $a_1$ (Fig.~\ref{fig:zro2-fitting}~(a)). For the two sets of structures obtained by freezing in $Q_{X^+_{5,y}}$, and separately $Q_{X^+_{5,x}}$=$Q_{X^+_{5,y}}$=$Q_{X^+_{5,z}}$, and relaxing, we obtain the energies and the relaxed values of $Q_{\Gamma_4^-}$. By fitting the relaxed values of $Q_{\Gamma_4^-}$ to the expression $Q_{\Gamma_4^-}$=$-$(a$_2$/2a$_1$)$Q_{X^+_{5,y}}$ obtained for both structures by minimizing Eq.~(1) with respect to $Q_{\Gamma_4^-}$, we extract the value of $a_2$ (Fig.~\ref{fig:zro2-fitting}~(b)). Finally, we use the computed energies of these two sets of structures to obtain $S(0,Q,0)$=$a_3Q^2$ and $S(Q,Q,Q)$ as a spline interpolation (Fig.~\ref{fig:zro2-fitting}~(c)). 
\medskip

\begin{figure}[t]
\includegraphics[scale=0.45]{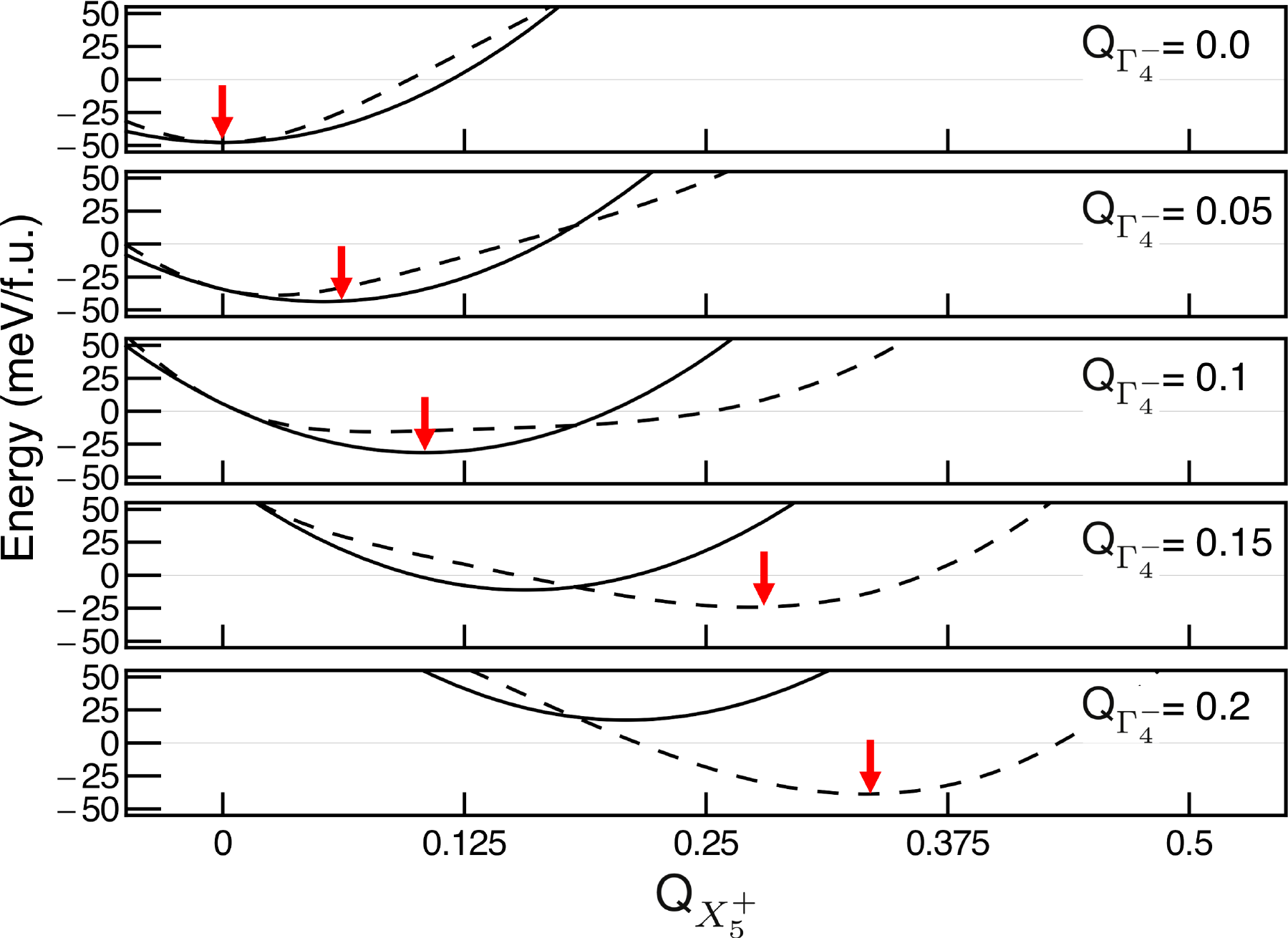}
\caption{\label{fig:zro2-model} Energy (E$_T$) as a function of Q$_{X_{5,y}^+}=Q_{X_5^+}$ (solid line) and Q$_{X_{5,x,y,z}^+}=Q_{X_5^+}$ (dashed line) at fixed values of the mode amplitude Q$_{\Gamma_4^-}$. The red arrows show the global energy minimum for each value of Q$_{\Gamma_4^-}$.}
\end{figure}

Using the model we can reproduce the energy cusp of Fig.~\ref{fig:zro2-barrier} by plotting 
\begin{equation*}
\min_{Q_{X_{5,x,y,z}^+}} E_T(Q_{\Gamma_4^-},Q_{X_{5,x,y,z}^+})
\end{equation*}
\noindent as a function of $Q_{\Gamma_4^-}$. As shown in Fig.~\ref{fig:zro2-barrier}, the model is able to capture the essential elements of the energy landscape: the position and height of the energy barrier, the two local minima and the energy difference between them. The disagreement between first principles and the model can be attributed to the truncation of the expansion in Eq.~(1). The cusp is caused by the multiple local minima displayed by the energy as a function of the X$_{5}^+$ mode, with a small change in the value of $Q_{\Gamma_4^-}$ causing a switch between two local minima (Fig.~\ref{fig:zro2-model}).
\medskip

We investigate the effect of epitaxial strain on the relative stability of the tetragonal, orthorhombic, and monoclinic phases. Epitaxial strain is simulated through ``strained-bulk" calculations~\cite{Pertsev1998, Dieguez2005}. We denote the epitaxially (e) strained \textsl{Pca2$_1$} structure as \textsl{ePca2$_1$}, and indicate the matching plane by a prefix labelling the normal to the plane (for example, normal vector $\texttt{t$_a$}$ yields \textsl{a-ePca2$_1$}).  
\medskip 

\begin{figure}[t]
\includegraphics[scale=0.35]{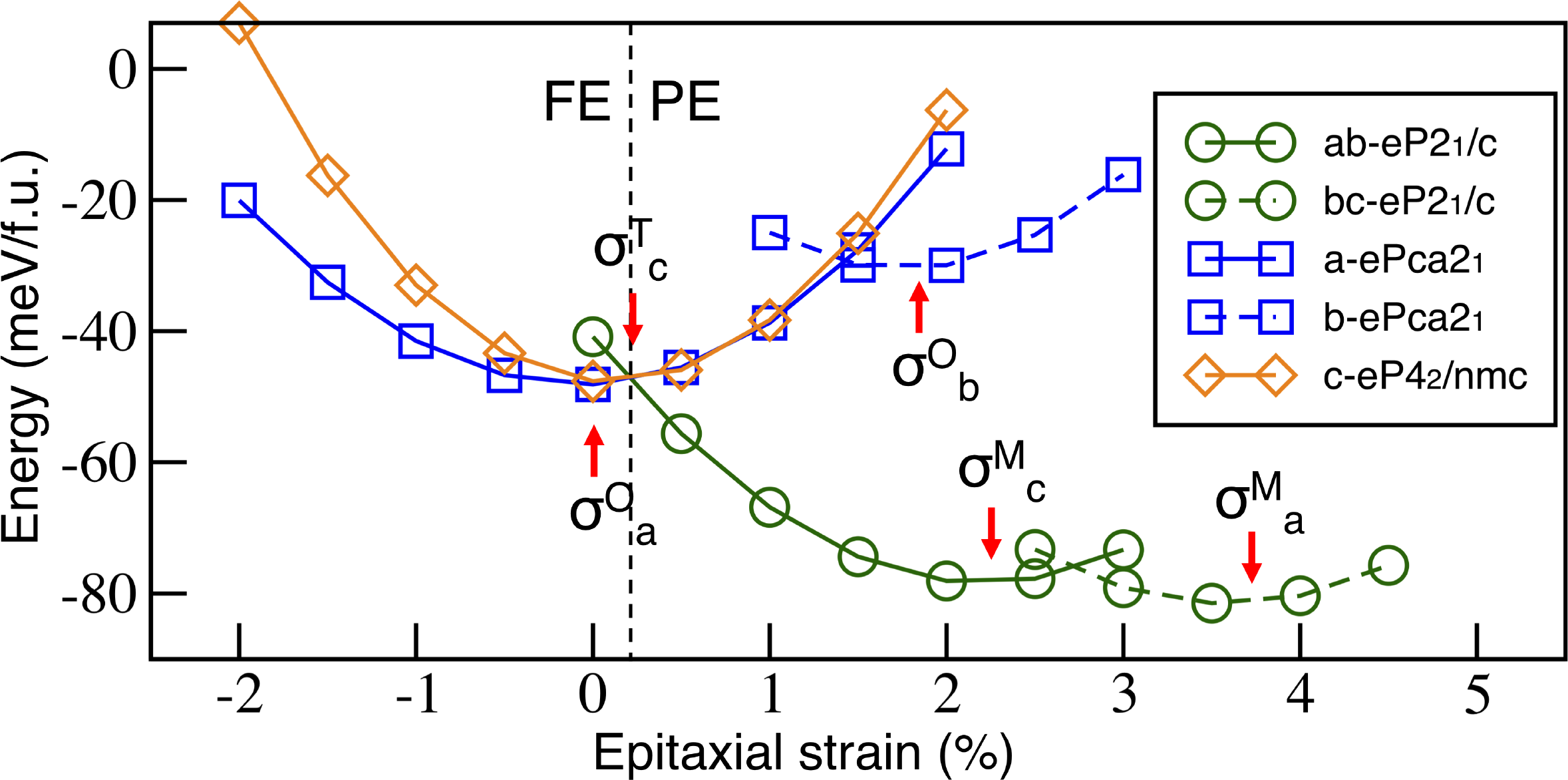}
\caption{\label{fig:zro2-epitaxial} Epitaxial strain diagram for ZrO$_2$. FE and PE refer to ferroelectric and paraelectric ground state.}
\end{figure}

Fig.~\ref{fig:zro2-epitaxial} shows the calculated epitaxial strain diagram for ZrO$_2$. As expected from the dissimilar lattice constants, the equilibrium energy strains ($\sigma$) for the three phases and various matching planes are quite different ($\sigma^M_{c}$=2.3\% and $\sigma^M_{a}$=3.8\% for monoclinic, $\sigma^O_{a}$=0.0\% and $\sigma^O_{b}$=1.9\% for orthorhombic, and $\sigma^T_{c}$=0.2\% for tetragonal phases, respectively). The computed values of $\sigma$ are in excellent agreement with the estimates obtained by comparing the relevant relaxed lattice vectors with the reference lattice constant a$_0$=5.03~\AA, as discussed in~\cite{Reyes-Lillo2013}.
\medskip 

For large values of tensile strain ($>$~$1$\%), the monoclinic phases are highly favorable in energy. As the strain is decreased, the relative stability of the monoclinic phase is reduced and the tetragonal and orthorhombic phases become favorable. In the strain range from $-1.0$\%~to~$+1.5$\%, the tetragonal and orthorhombic phases are very close in energy. In this regime, if the appearance of the lower-energy monoclinic phase is suppressed by surface effects~\cite{Garvie1978, Kim2004, Keun2008} or other means~\cite{Fischer2008a,Fischer2008b}, the system is expected to be antiferroelectric. For large values of compressive strain ($<$~$-1$\%), the ferroelectric structure \textsl{a-ePca2$_1$} is favored. While within the accuracy of our calculations it is not possible precisely to predict the critical strains that will be observed in experiments, we expect stabilization of ferroelectricity at accessible values of compressive strain. 
\medskip 

Finally, as an illuminating comparison with ZrO$_2$, we consider the stability of the corresponding tetragonal, orthorhombic, and monoclinic structures in HfO$_2$. The calculated lattice constants for the structures of HfO$_2$ are shown in Table~\ref{table:polar-phases-hfo2}. The smaller ion size of Hf$^{+4}$ compared to Zr$^{+4}$ explains the reduced lattice constants of HfO$_2$, in good agreement with previous calculations~\cite{Zhao2002b}. The computed spontaneous polarization of~60~$\mu$C/cm$^2$ is consistent with previous results~\cite{Clima2014}. The main difference between ZrO$_2$ and HfO$_2$ is the higher relative energy of the tetragonal structure in HfO$_2$, so that the polar orthorhombic structure is favored over the nonpolar tetragonal phase by~$\sim$~23~meV/f.u. This suggests that isovalent substitution of Zr by Hf will favor the polar orthorhombic phase in thin films of (Zr$_{1-x}$Hf$_{x}$)O$_2$, consistent with Refs.~\onlinecite{Muller2012, Muller2011b, Park2013}.
\medskip

Estimation of the minimum energy strains for the relevant structures of HfO$_2$ gives similar values to the case of ZrO$_2$ ($\sigma^M_{c}$=2.4\% and~$\sigma^M_{a}$=3.7\% for monoclinic, $\sigma^O_a$=$-0.1$\% and $\sigma^O_b$=$-1.0$\% for orthorhombic, and $\sigma^T_c$=0.2\% for tetragonal structures, respectively). Therefore, the epitaxial strain diagram should be similar to that of ZrO$_2$ (Fig.~\ref{fig:zro2-epitaxial}) but with the epitaxial strain curve of the tetragonal structure shifted higher by~$\sim$~23~meV/f.u. Based on this, we speculate that ferroelectricity would be observed in HfO$_2$ over a wide range of epitaxial strain if the monoclinic ground state were suppressed.   
\medskip

\begin{table}[t]
\caption{Energy $\Delta E$ (meV/f.u.), lattice constants (\AA), monoclinic angle ($\beta$), and volume expansion $\Delta V/V$ ($\%$) for the ZrO$_2$-type phases of HfO$_2$.}
\begin{ruledtabular}
\begin{tabular}{lcccccc}
Phase  & $\Delta E$ & a & b & c & $\beta$ & $\Delta V/V$   \\ \hline
\begin{tabular}{l} Cubic\\ Tetragonal\\ Orthorhombic\\ Monoclinic\\ \end{tabular}
&\begin{tabular}{c} 0\\ -22.5\\ -45.1\\ -93.2\\ \end{tabular}
&\begin{tabular}{c} 4.89\\ 4.90\\  5.07\\ 4.95\\ \end{tabular}
&\begin{tabular}{c}     \\     \\  4.88\\ 5.06 \\ \end{tabular}
&\begin{tabular}{c}     \\ 4.95\\  4.89\\ 5.08 \\ \end{tabular}
&\begin{tabular}{c}     \\     \\      \\ 99.53\\ \end{tabular}
&\begin{tabular}{c} 0\\ 1.3\\ 3.4\\ 7.0\\ \end{tabular} \\ 
\end{tabular}
\end{ruledtabular}
\label{table:polar-phases-hfo2}
\end{table}

In summary, the experimentally observed field-induced ferroelectric transition corresponds to an intrinsic behavior of thin-film ZrO$_2$. The tetragonal-to-orthorhombic antiferroelectric transition is explained as a field-induced first-order transition where the polar mode is stabilized from the cubic phase by a coupling with two zone-boundary modes. Our results suggest that a ferroelectric phase could be favored over the antiferroelectric phase by appropriate epitaxial strain or isovalent substitution of Zr for Hf.  The absence of toxic elements and the compatibility of ZrO$_2$ with silicon make these results especially relevant for technological applications.
\medskip

\section{\label{sec:level1} Acknowledgments}

We thank S. V. Kalinin, T. Mikolajick, J. M$\ddot{u}$ller, T. Schenk, D. G. Schlom, U. Schr$\ddot{o}$der, and D. Vanderbilt for valuable discussions. S.E.R.-L. would like to thank S. Trolier-McKinstry and C. Randall for their hospitality at MRI-Penn State. This work was supported by the Office of Naval Research Grant No.~N00014-12-1-1040. S.E.R.-L. would also like to thank the support of Conicyt and the sponsor of Fulbright Foundation.


\begin{thebibliography}{99}
\bibitem{fm-3m-irrep} Note that each of these modes transform as a single partner function of a multidimensional irrep of \textsl{Fm$\bar{3}$m}.
\bibitem{pca21-variant} We note the fact that freezing $X_{5,x}^+$ or $X_{5,z}^+$ alone in the tetragonal structure gives rise to a \textsl{Pca2$_1$} structure distinct from the orthorhombic phase of ZrO$_2$ considered here.
\end{thebibliography}
\end{document}